# Encapsulation and Electronic Control of Epitaxial Graphene by Photosensitive Polymers and UV light.[1]

Samuel Lara-Avila[1], Kasper Moth-Poulsen[2], Rositza Yakimova[3], Thomas Bjørnholm[4], Vladimir Fal'ko[5], Alexander Tzalenchuk[6], and Sergey Kubatkin[1]

[1]Department of Microtechnology and Nanoscience, Chalmers University of Technology, Göteborg, S-41296 (Sweden)

[2]College of Chemistry, University of California, Berkeley, CA 94720 (USA)

[3]Department of Physics, Chemistry and Biology (IFM). Linköping University, Linköping, S-581 83 , (Sweden)

[4]Nano-Science Center & Department of Chemistry, University of Copenhagen, Copenhagen, DK-2100 ø (Denmark)

[5]Physics Department, Lancaster University, Lancaster, LA1 4YB (UK)

[6]National Physical Laboratory, Teddington, TW11 0LW (UK)

Dated: (January 31, 2011)

Electronic devices using epitaxial graphene on Silicon Carbide require encapsulation to avoid uncontrolled doping by impurities deposited in ambient conditions. Additionally, interaction of the graphene monolayer with the substrate causes relatively high level of electron doping in this material, which is rather difficult to change by electrostatic gating alone.

Here we describe one solution to these problems, allowing both encapsulation and control of the carrier concentration in a wide range. We describe a novel heterostructure based on epitaxial graphene grown on silicon carbide combined with two polymers: a neutral spacer and a photoactive layer that provides potent electron acceptors under UV light exposure. Unexposed, the same double layer of polymers works well as capping material, improving the temporal stability and uniformity of the doping level of the sample. By UV exposure of this heterostructure we controlled electrical parameters of graphene in a non-invasive, non-volatile, and reversible way, changing the carrier concentration by a factor of 50. The electronic properties of the exposed SiC/ graphene/polymer heterostructures remained stable over many days at room temperature, but heating the polymers above the glass transition reversed the effect of light.

The newly developed photochemical gating has already helped us to improve the robustness (large range of quantizing magnetic field, substantially higher opera- tion temperature and significantly enhanced signal-to-noise ratio due to significantly increased breakdown current) of a graphene resistance standard to

such a level that it starts to compete favorably with mature semiconductor heterostructure standards. [2,3]